\documentclass[8pt]{iopart}

\usepackage{epsfig}
\usepackage{graphicx}

\newcommand{\beq}{\begin{equation}}
\newcommand{\eeq}{\end{equation}}

\newcommand{\vecP}{{\bf P}}

\newcommand{\vecE}{{\bf E}}

\newcommand{\veck}{{\bf k}}

\newcommand{\vecn}{{\bf n}}
\newcommand{\vece}{{\bf e}}
\newcommand{\cale}{\mbox{\boldmath{$\mathcal{E}$}}}
\begin{document}

\title{Landau damping in thin films irradiated by a strong laser field}

\author{D~ F~Zaretsky$^{1,3}$, Ph~A~Korneev$^{2*}$,
\footnote[0]{$^*$ e-mail: korneev@theor.mephi.ru}
S~V~Popruzhenko$^{2,3}$ and W~Becker$^3$}

\address{$^1$  Russian Research Center ``Kurchatov Institute",
123182, Moscow, Russia}

\address{$^2$ Department of Theoretical Nuclear Physics,
Moscow Engineering Physics Institute, Moscow, 115409, Russia}

\address{$^3$ Max-Born-Institut, 12489, Berlin, Germany}

\begin{abstract}
The rate of linear collisionless damping (Landau damping) in a classical
electron gas confined to a heated ionized thin film is calculated.
The general expression for the imaginary part
of the dielectric tensor in terms of the parameters of the single-particle self-consistent 
electron potential is obtained.
For the case of a deep rectangular well, it is explicitly calculated as a function
of  the 
electron temperature in the two limiting cases of specular and diffuse reflection 
of the electrons from the boundary of the self-consistent potential.
For realistic experimental parameters, the contribution of  
Landau damping to the heating of the electron subsystem is estimated. 
It is shown that for films with a thickness below about 100 nm and for 
moderate laser intensities it may be comparable with or even dominate 
over electron-ion collisions and inner ionization.

\end{abstract}

\submitto{\jpb}

\maketitle

\section{Introduction}
The interaction of strong laser fields with  nanobodies such as clusters and
thin films has become a subject of wide interest over the last decade.
A very significant distinguishing feature of clusters and thin films 
irradiated by a strong laser field as compared with a gas of isolated atoms and even with
solids and plasmas is 
the very efficient  energy transfer per atom
from the laser field to the respective matter \cite{dm96}.
The high efficiency of energy absorption in nanobodies produces fast heating of
their electron subsystem up to temperatures on the keV scale
and leads to the occurrence of many interesting new phenomena
such as Coulomb explosion of clusters, production of ions with very high
charges, x-ray emission from cluster targets and films, to name a few.
Therefore, the mechanisms of energy absorption by the electron subsystem 
in nanobodies are of primary interest and importance for understanding 
many aspects in strong-field nanophysics.

The physics of laser-irradiated nanostructures including clusters and
thin films have been reviewed recently \cite{gord,calvayrack,krainov}. 
Various models of the  energy absorption from a laser pulse were
discussed.
In the nanoplasma model \cite{dm96,dm00}, which is meant to describe clusters
irradiated by a strong laser field with intensities of $10^{15}$--$10^{18}$ W/cm$^2$,
the imaginary part of the dielectric constant  
$\varepsilon=\varepsilon^{\prime}+\rmi \varepsilon^{\prime\prime}$ is supposed 
to be the result of individual electron-ion collisions like in 
an infinite plasma, so that $\varepsilon^{\prime\prime}=2\nu_{\rm ei}/\omega$.
Here $\nu_{\rm ei}$ is the frequency of electron-ion collisions and $\omega$ the laser frequency.
In a plasma with parameters typical for intense-laser--cluster interaction, 
the time between individual collisions is around 10--100 fs.
For laser pulses of moderate intensity ($10^{13}$--$10^{14}$ W/cm$^2$), as they are 
used in studies of the optical breakdown in dielectrics 
(see \cite{rudolph,kaiser,rudolph2} and references therein),  
electron-phonon collisions give a relaxation time on the picosecond scale.

Another efficient mechanism of energy absorption is vacuum heating
\cite{brunel}.
However, it is relevant only for a not too thin plasma and 
for high laser intensities when the conditions $a>x_0>l$
are satisfied, where $a$ is the thickness of the plasma, $x_0$ the
amplitude of the electron oscillations under the action of the laser field, and
$l$ the size of the skin layer.
For the parameters considered here, namely films with a thickness 
$a\le 100$ nm and moderate laser intensities, the relation between $a,x_0$ and
$l$ is rather the other way around, so that the mechanism of vacuum heating is irrelevant.

Finally, laser energy may be absorbed in electron collisions
with the confining surface.
This effect can be interpreted as  collisionless damping, or Landau
damping, in a spatially restricted plasma.
Such a mechanism of energy transfer  in various multiparticle systems
from collective into individual-particle degrees of freedom was first considered  
 for
longitudinal waves in a quasi-neutral infinite plasma \cite{dau46,dau9}. 
For a finite system, Landau damping was studied for the
first time in 1966 \cite{kubo} for a collection of cold fine metallic particles, assuming 
an external field weak enough to leave the electron 
gas at zero temperature. 
In a cold spherical nanoplasma, the imaginary part of the dielectric constant 
produced by Landau damping has the form
\begin{equation}
\varepsilon^{\prime\prime}\equiv\varepsilon_{\rm L}(\omega)=C(\omega) {v_{\rm F}\over\omega a},
\label{cold}
\end{equation}
where $v_{\rm F}$ is the Fermi velocity, $a$  the spatial size of the system
(\textit{e.g.} $a=R$ for a spherical cluster), and the dimensionless function $C(\omega)$
depends on the boundary conditions for the electrons and on 
the geometry of the system.
A detailed discussion of Landau damping in cold metal clusters
including an analysis of the role of boundary conditions for the electrons
and a comparison with experimental data may be found in references 
\cite{hache,kreibig,fomichev}.
For cold particles with radii below 100 nm, collisionless damping
dominates over collisions with phonons \cite{kreibig}, viz. 
$\varepsilon_{\rm L}\gg\varepsilon_{\infty}$,
so that the absorptive part (\ref{cold}) determines the cross 
sections of light scattering 
and absorption by a small nanobody with a cold degenerate electron subsystem.

For nanostructures interacting with a strong laser field, the state of the  
electron gas differs significantly from that in cold clusters:
first, the  electron gas is heated to 
a temperature on the keV scale, so that the electron plasma becomes 
purely classical; second, 
 a nanobody irradiated by a strong field becomes partially 
ionized so that the self-consistent electron potential 
may differ considerably from the rectangular well
used for the description of cold particles \cite{kubo,hache,fomichev}. 
However, the contribution of Landau damping to absorption remains
substantial \cite{kost1,megi,kost2}.
In particular, it defines the amplitude of the electric field inside a cluster 
near the Mie resonance \cite{megi}.
The latter may become particularly important, when it is \textit{nonlinearly} 
excited. This was recently proposed theoretically \cite{fom03,fom04} and observed experimentally 
\cite{hays} in laser-irradiated clusters.
In references \cite{megi,fom03}, the width of the Mie resonance was estimated assuming
that the frequency of electron collisions with the boundary is 
\beq
\nu_{\rm L}\simeq{v\over R},
\label{nuL}
\eeq
where $v$ is the thermal or effective electron velocity.
This crude estimate yields for the  imaginary part of the dielectric constant
the expression (\ref{cold}) with the effective velocity $v$ in place of 
$v_{\rm F}$\footnote{In \cite{megi}, the effective velocity was
taken in the form $v_{\rm eff}=\sqrt{v^2_{\rm T}+v^2_{\rm q}}$, where $v_{\rm T}$ and $v_{\rm q}$
are the thermal and quiver velocities, respectively.}
and $C=2$.
Obviously, it does not take into account the shape of the self-consistent
potential whereby the electrons are trapped nor the shape of the thermal distribution
function.
Therefore, the effect of Landau damping in a heated classical nanoplasma
is still lacking a quantitative treatment.

In the present work, we calculate for the first time
the contribution to the dielectric tensor from collisionless damping 
in the heated electron plasma trapped inside an ionized laser-irradiated
one-dimensional (1D) thin film.
The plasma is supposed to be classical, 
so that this contribution does not depend on Planck's constant.
As we shall show, in a classical finite system the simple form (\ref{nuL}) is only realized 
under specific conditions. In general, the behavior of the imaginary part $\varepsilon_{\rm L}$ of 
the dielectric tensor is quite complicated and depends on the shape of the self-consistent
single-particle electron potential and on the specific boundary conditions.
Its numerical value can be substantial, so that this mechanism
may be dominant in the process of electron plasma heating.
Some preliminary results of this work were annonced in \cite{lphys}.

The paper is organized as follows.
In the next section, we state the problem and obtain 
the general expressions for the rate of collisionless energy absorption
and for the dielectric tensor $\varepsilon_{ik}(\omega)$
in a homogeneous isotropic thin film.
The case of diffuse reflection of electrons from the boundary of the  
self-consistent potential is considered in the third section.
The fourth section contains the application of the results obtained to the 
description of laser light absorption in a thin film. 
Details of the calculations are delegated to appendixes A and B.

\section{Statement of the problem and general expression for the 
dielectric tensor in a thin film}

Let us consider a one-dimensional system (a thin film) consisting of ions
homogeneously distributed inside a layer of width $a$ and a heated electron
plasma concentrated mainly within this ion layer.
As a result of the partial evaporation of the electrons, the system has some 
nonzero positive charge and produces  a 
self-consistent static electric potential $U(z)$, which traps the residual 
electrons.
We treat the ionic subsystem as frozen, which usually is a valid assumption 
on the femtosecond time scale\footnote{For clusters irradiated
by a strong laser field,  the time when this assumption ceases to be valid 
has been estimated, for example, in 
reference \cite{last}. 
This time can be measured experimentally via  the
dependence of the intensity of the scattered light on the laser pulse duration
\cite{dm00}. Both the estimates and the experimental results support the conclusion
that for nanobodies with radii around 100 A the approximation of
frozen ions holds up to a few hundred 
femtoseconds.}.
The self-consistent potential obeys the Poisson equation

\beq
\Delta U(z)=-4\pi e[z_{\rm i}n^0_{\rm i}(z)-n^0_{\rm e}(z)],
\label{poisson}
\eeq
where $e$ is the absolute value of the electron charge,
$n^0_{\rm i}(z)$ and $n^0_{\rm e}(z)$ are the equilibrium
concentrations of ions and electrons, respectively. Inside the ion layer, 
i.e. for $-a/2\le z\le a/2$, we have $n^0_{\rm i}(z)=n_0$.  
Equation (\ref{poisson}) is not sufficient to define $U(z)$, because the 
equilibrium electron distribution $n^0_{\rm e}(z)$ is unknown, too.
For the consistent definition of these two functions,  equation (\ref{poisson}) 
must be solved simultaneously with the Boltzmann equation for an electron plasma.
Here, we do not consider this problem and take 
the self-consistent potential as a given function.

Oscillations of the electron cloud in such a system arise 
either due to the action of the external laser field or due to
a small deviation of the electron cloud from its equilibrium position.
We suppose that the rate of electron-electron and electron-ion collisions 
is low, so that oscillations of the electron cloud are far off  
equilibrium and
do not change the electron distribution in momentum space.
The oscillations produce the perturbation $\delta n_{\rm e}(x,t)$ of the spatial electron
distribution.
If their amplitude is small
with respect to the spatial size $a$ of the layer, 
the oscillations may be considered as homogeneous (dipole) 
with the laser frequency $\omega$\footnote{Free
oscillations take place  at the plasma frequency 
$\omega_{\rm p}=\sqrt{4\pi e^2n_0\kappa/m_{\rm e}}$,
where $m_{\rm e}$ is the electron mass and 
$\kappa$ is the fraction of electrons trapped inside the ion layer.}.
This approximation corresponds to the model of an incompressible
fluid for the electrons.
For clusters in strong laser fields, this model was used in reference \cite{fom03}, 
and its applicability was supported by numerical simulations
\cite{fom04}.
The time-dependent electron density then is given by a rigid displacement by $\xi(t)$
from its equilibrium density  $n_{\rm e}^0(z)$: 

\beq
n_{\rm e}(z,t)\equiv n^0_{\rm e}(z)+\delta n_{\rm e}(z,t)=n^0_{\rm e}(z-\xi(t)),
\label{dipole}
\eeq
where the displacement obeys the nonlinear oscillator 
equation\footnote{For a spherical
cluster, such equations are derived in references \cite{parks,fom03,rost}} 
\begin{eqnarray}
\ddot\xi+2\gamma(\omega)\dot\xi+\omega_{\rm p}^2F(\xi)=-{eE_{0z}\over m_{\rm e}}\cos(\omega t+\alpha),
\label{osc}\\
F(\xi)=\xi\bigg\lbrack 1+\sum_{k=1}^{\infty}b_k\bigg({\xi\over a}\bigg)^{2k}\bigg\rbrack,
\label{xi}
\end{eqnarray}
where the external force is provided by the laser field\footnote{Of course, the phase $\alpha$
has no physical relevance at this point. 
It could be removed by just resetting time. However,
it will be convenient in the treatment of diffuse reflection below.} 

\beq
\vecE(t)=\vecE_0\cos(\omega t+\alpha),
\label{laser}
\eeq
and we define the $z$ axis perpendicular to the plane of the film (see Fig. 1).
The damping constant $\gamma$ in equation (\ref{osc}) is related to the
imaginary part of the dielectric tensor [see equation (\ref{gamma}) below]. 

Here we consider  linear damping.  Therefore, it is enough to 
retain the first term 
in the expansion of $F$ in equation (\ref{xi}). 
The resulting harmonic oscillator has the solution

$$
\xi(t)=-{eE_{0z}\over m_{\rm e}\sqrt{(\omega_{\rm p}^2-\omega^2)^2+
4\gamma^2\omega^2}}\cos(\omega t+\alpha+\beta),
\nonumber
$$

\beq
\sin\beta=-{2\gamma\omega\over\sqrt{(\omega_{\rm p}^2-\omega^2)^2+
4\gamma^2\omega^2}}.
\label{xisol}
\eeq

The electric field inside the thin film is the sum of the external field and the induced
field.
Assuming $\gamma\ll\omega$ we obtain 

$$
\vecE(\xi,t)=4\pi en_0\xi\vece_z+\vecE_0\cos(\omega t+\alpha)\equiv
\cale_{0\bot}\cos(\omega t+\alpha)+\vece_z{\mathcal{E}}_z\cos(\omega t+\alpha-\delta),
\nonumber
$$
\beq
\cale_{0\bot}=\vecE_{0\bot},~~~
{\mathcal{E}}_z={E_{0z}\over{\sqrt{(1-\omega_{\rm p}^2/\omega^2)^2+
4\gamma^2/\omega^2}}},~~~
\sin\delta=\bigg({\omega_p\over\omega}\bigg)^2\sin\beta.
\label{E}
\eeq

The rate $\overline{Q}$ of collisionless energy absorption is equal 
to the work per unit time of the electric field (\ref{E}) on the individual 
electrons averaged over their distribution.
For a classical system as we consider it here, it is independent of  Planck's  
constant. Yet, surprisingly, the simplest route  for its calculation
is via quantum mechanics. 
Following \cite{callen}, we consider electrons populating energy levels 
in the symmetric potential well
$U(z)=U(-z)$, with some distribution function $\rho(\epsilon)$ and
subject to  the uniform electric field (\ref{E}).
The electron energy $\epsilon$ depends only on the principal quantum number $n$. 
The field produces electron transitions from their initial level $n$ to a final level 
$n_1$. 
The rate of energy absorption from the laser field per unit time and per unit square of the film is  
\beq
\overline{Q}=\hbar\omega n_{\rm e}a\sum_{n,n_1}
\rho(\epsilon_n)(w^-_{n,n_1}-w^+_{n,n_1}),
\label{Q1}
\eeq
where $w^-_{n,n_1}$ and $w^+_{n,n_1}$ are the transition probabilities per unit time  of an
electron from the level $n$ to the level  $n_1$ owing to 
absorption or emission of a laser photon, respectively.
The distribution function $\rho$ may also depend on time, due to 
evaporation of electrons from the well. 
Moreover, individual collisions or fluctuations of the self-consistent 
potential may influence the energy absorption.
However, here we restrict ourselves to the simplest limiting case, 
when individual collisions and fluctuations may be neglected and the 
quantum state of the electrons may be described in terms of wave functions.
Classically, such an approximation corresponds to the motion 
of electrons in a given time-independent potential with specular reflection
from the boundary whose position depends on the electron energy.
The opposite limiting case of diffuse reflection of the electrons from the
boundary will be briefly considered  in the next section.

We calculate the  probabilities $w^{\mp}_{n,n_1}$ in  
first-order perturbation theory with respect to the field (\ref{E}).
Although the external laser field is strong, far from the plasma resonance 
the amplitude of electron cloud oscillations satisfies 
$x_0\approx eE_{0z}/m_{\rm e}\omega^2_{\rm p}\ll a$.
So, for a laser pulse with intensity $I\simeq 10^{15}$ W/cm$^2$, 
a film with thickness $a\simeq 10$ nm, and the electron density 
$n_{\rm e}\simeq 10^{23}$ cm$^{-3}$,  we have $\omega_{\rm p}\approx 8$ eV
and $x_0\approx 1$ nm.
These estimates support the applicability of lowest-order perturbation theory.
The details of the calculation are given in Apendix A.
Finally, then, the rate of energy absorption has the form

\beq
\overline{Q}=-{\pi n_{\rm e}ae^2{\mathcal{E}}_z^2\over 2m_{\rm e}^2\omega}\sum_{k=0}^{\infty}
{\vert p_{2k+1}(\epsilon_k)\vert^2\over
\vert\Omega^{\prime}_{\epsilon}(\epsilon_k)\vert}
{\rmd\over \rmd\epsilon}\bigg\lbrack f(\epsilon)\Omega(\epsilon)\bigg\rbrack
_{\epsilon=\epsilon_k}.
\label{main_1d}
\eeq
Here the sum extends over all roots of the equation

\beq
(2k+1)\Omega(\epsilon_k)=\omega,
\label{omega}
\eeq
where $\Omega(\epsilon)$ is the round-trip frequency of an electron with energy
$\epsilon$ in the self-consistent potential $U(z)$. The condition (\ref{omega}) 
admits a simple and transparent interpretation: a nonzero 
contribution to energy absorption comes from those 
electrons only whose period of motion in the potential $U(z)$ is equal to an 
odd number of laser periods.
Therefore, as in the case of an infinite plasma or cold finite plasma, 
energy absorption proceeds through ``resonant'' particles.

The expression for the dielectric tensor $\varepsilon_{ik}(\omega)$ may be
found by a comparison of the result (\ref{main_1d}) with the standard
formula for the rate of energy absorption from a plane electromagnetic wave 
interacting with a dielectric \cite{dau8},

\beq
\overline{Q}=-{1\over 2}{\rm Re}\,\vecP\cdot\dot{\vecE}^*,
\label{QPE}
\eeq
where $P_i=(\varepsilon_{ik}-\delta_{ik}){\mathcal{E}}_k/4\pi$ is the polarization per 
unit volume, $\cale$ is the electric field (\ref{E}) inside the film,
and $\vecE$ the incident field (\ref{laser}).
For a homogeneous isotropic film, the dielectric tensor should be proportional 
to the unit tensor $\delta_{ik}$.
However, Landau damping produces anisotropy since only 
oscillations along the $z$ axis lead to energy absorption.
As a result, the dielectric tensor in a homogeneous isotropic film 
has the form
\beq
\varepsilon_{ik}(\omega)=\bigg(\varepsilon_0(\omega)
+1-{\omega_{\rm p}^2\over\omega^2}\bigg)
\delta_{ik}+\rmi \varepsilon_{\rm L}(\omega)\delta_{iz}\delta_{kz},
\label{tensor}
\eeq
where $\varepsilon_0=\varepsilon_0^{\prime}+\rmi\varepsilon_0^{\prime\prime}$
is the contribution from the polarizability of the bound electrons and 
electron-ion collisions.
A nonlinear contribution to $\varepsilon_0^{\prime\prime}$ due to 
multiphoton ionization can also be included.
The anisotropic term $\varepsilon_{\rm L}$ is responsible for Landau damping.

Expanding the external field into p-polarized and s-polarized waves 
as shown in Fig. 1, we get from equations (\ref{QPE}) and (\ref{tensor}) (\textit{cf.} figure 1 for the definition of $s$ and $p$)
\beq
\overline{Q}={a\omega\over 8\pi}E_0^2\bigg\lbrack 
{\varepsilon_{\rm L}+\varepsilon_0^{\prime\prime}\over\vert\varepsilon\vert^2}p^2
\sin^2\vartheta+\varepsilon_0^{\prime\prime}(s^2+p^2\cos^2\vartheta)
\bigg\rbrack.
\label{Q3}
\eeq
Comparing equations (\ref{main_1d}), (\ref{Q3}), and (\ref{E}) we find
\beq
\gamma(\omega)={\varepsilon^{\prime\prime}_{zz}(\omega)\over 2\omega}=
{\varepsilon_{\rm L}+\varepsilon_0^{\prime\prime}\over 2\omega},
\label{gamma}
\eeq

\beq
\varepsilon_{\rm L}(\omega)=-{\pi\over m_{\rm e}}\bigg({\omega_{\rm p}\over\omega}\bigg)^2
\sum_{k=0}^{\infty}{\vert p_{2k+1}(\epsilon_k)\vert^2\over
\vert\Omega^{\prime}_{\epsilon}(\epsilon_k)\vert}
{\rmd\over \rmd\epsilon}\bigg\lbrack f(\epsilon)\Omega(\epsilon)
\bigg\rbrack_{\epsilon=\epsilon_k}.
\label{epsL}
\eeq

Expression (\ref{epsL}) is the main result of the present work.
It describes the contribution of Landau damping to 
the dielectric tensor for a homogeneous isotropic thin film.

\section{Dependence on the boundary conditions}
In our calculations so far, we have assumed that an electron moves inside
a thin film under the action of the field (\ref{E}) and the self-consistent single-particle 
potential $U$ produced by the ion core and the quasi-equilibrium distribution of
all electrons.
Individual collisions and fluctuations both have been completely ignored.
In macroscopic language, this approximation corresponds to the specular 
reflection of an electron from the boundary of a self-consistent potential.
Below, we identify the dielectric constant calculated in this approximation by
the superscript ``s'', $\varepsilon_{\rm L}\to\varepsilon_{\rm L}^s$.
In this case, the influence of the electric field (\ref{E})
on the electron motion is a \textit{coherent} effect accumulated over many periods.
This leads to the feature expressed by equation (\ref{omega}):
the energy absorption from the electric field is possible
only for selected electrons with ``resonant'' energies $\epsilon_k$,
as explained below equation (\ref{omega}).
All other energy levels do not contribute to $\varepsilon_{\rm L}^s$.
However, the approximation of specular reflection is only a limiting case,
since collisions are present and do have an influence on the efficiency  of energy transfer.
The reason is that they interrupt the coherent accumulation of the phase of the electron motion.
In consequence, energy absorption is possible even if the resonance condition
(\ref{omega}) is not satisfied.
This may result in increased  absorption 
in comparison with  the case of specular reflection.

A rigorous quantum-mechanical account of collisions requires  a 
density-matrix approach.
However, an estimate of the effect of collisions on Landau damping
may be obtained using a simple classical method.
The influence of individual collisions on the electron trajectory
in a self-consistent field increases when the velocity of the 
electron is small, i.e. in the vicinity of the turning points.
Here, an electron can experience a large momentum transfer in an individual collision
and, as a result, will lose the memory of its phase.
To estimate the effect of this phase mismatch, we use a simple model for 
diffuse reflection of an electron on the boundary.
We suppose, that \textit{between the boundaries} 
every electron moves in a self-consistent potential under the action of the
field (\ref{E}), but \textit{in the collision with the boundary} the phase is reset, 
 so that subsequent trips of the electron between the boundaries
 are independent.
In this approximation, the total energy that the electron collects from the field
(\ref{E}) is given by the \textit{incoherent} sum of the contributions from the various
segments of the electron motion between the turning points.
Treating the electron classically, instead of (\ref{Q1}) we have for the rate of energy absorption
by the 1D system 

\beq
Q=\lim_{\tau\to\infty}{1\over\tau}\int \rmd\epsilon f(\epsilon)A(\epsilon,\tau),
\label{diffQ}
\eeq
where $A(\epsilon,\tau)$ is the energy absorbed during the time $\tau$
by the electron with the energy $\epsilon$ averaged over the initial 
conditions.
Obviously, this energy is the incoherent sum of the contributions from the various  
halfperiods within the time $\tau$ [see equation (\ref{diffQe}) below].

In the linear approximation used in this paper, 
averaging over the initial conditions is equivalent with averaging 
over the phase $\alpha$ of the field (\ref{E}) at the instant of time when
the electron starts from the boundary, so that 

\beq
A(\epsilon,\tau)={N(\epsilon,\tau)\over 2\pi}\int\limits_0^{2\pi}\rmd\alpha
\int\limits_0^{\tau(\epsilon,\alpha)}
\rmd tv_z(\epsilon,\alpha,t){\mathcal{E}}_z\cos(\omega t+\alpha),
\label{diffQe}
\eeq
where $\tau(\epsilon,\alpha)$ is the time the electron  
with energy $\epsilon$ needs to travel across the well from one boundary to the other, 
provided it started its motion at $t=0$ when the electric field 
has the phase $\alpha$.
This time differs from a half period of the unperturbed motion, viz. 
$\pi/\Omega(\epsilon)$,
by terms that are linear in the amplitude of the field.
The quantity $N$ is the number of half periods within the time $\tau$. so that
$N=\tau/(\pi/\Omega(\epsilon))$. 

The velocity $v(\epsilon,t)$ of the electron is to be found from the 
equation 
\beq
\dot v_z=-{1\over m_{\rm e}}{\partial U\over\partial z}
-{e\over m_{\rm e}}\tilde{E}_z\cos(\omega t+\alpha)
\label{diffv}
\eeq
with initial conditions $z(0)=z_0(\epsilon)$, where $z_0(\epsilon)$ 
is one of the turning points for the motion at the energy $\epsilon$.

In contrast to the case of specular reflection when the general expression 
(\ref{main_1d}) for the rate of energy absorption is, in principle, a  
functional of the self-consistent potential, equation (\ref{diffv}) 
cannot be solved in closed form for an 
arbitrary potential $U(z)$ even in linear approximation with respect to
the field (\ref{E})\footnote{This means that the solution of equation (\ref{diffv})
cannot be analytically expressed via the solution of the unperturbed equation
as was done in the previous section for specuilar reflection. 
In linear approximation with respect to the field (\ref{E}), 
equation (\ref{diffv}) reduces to the Hill equation.}.
Therefore, the case of diffuse reflection can be considered analytically
only for some particular self-consistent potential.
In the next section, we will consider and compare the results for specular 
and diffusive reflection for the simplest case of an infinitely deep rectangular well.

\section{Discussion}
If the distribution function $f(\epsilon)$ of the electrons and
the self-consistent potential $U(z)$ are known, the
contribution to the dielectric constant (\ref{epsL}) and
all relevant quantities such as the reflection and absorption coefficients
and the rate of energy absorption can easily be calculated
from equations (\ref{main_1d}), (\ref{tensor}), and (\ref{epsL}), either numerically or even 
in analytical form.
A separate and still unsolved problem is the determination of 
the quasi-equilibrium distribution function and the self-consistent
potential for electrons in a heated ionized film.
Here we consider the arguably simplest model: the infinitely deep rectangular 1D
well, which permits a complete analytical treatment both for 
specular and for diffuse boundary conditions.
This is an example  of a `closed' potential, 
where electrons cannot leave the system even for arbitrarily high energy.
In other words, there is no evaporation of electrons from the film after the
self-consistent potential has been established.
In such a potential, global thermodynamic equilibrium will be achieved
and electrons will be distributed according to the Gibbs law, so that
\beq
f(\epsilon)={\exp(-\epsilon/T)\over{\Omega(\epsilon)Z_0(T)}},
~~~~~~
Z_0(T)=\int {\rmd\epsilon\over\Omega(\epsilon)}\exp(-\epsilon/T).
\label{Z}
\eeq
This model provides a simple analytic expression for the dielectric constant,
which allows one to understand general properties of collisionless
absorption in a classical finite system.

Details of the calculations are given in Appendix B.
Both for specular and diffuse boundary conditions
the dielectric constant $\varepsilon_{\rm L}$ is the product of
the factor $\omega_{\rm p}^2/\omega^2$ and a function of the single parameter 
$x=\sqrt{8T/\pi m_{\rm e}\omega_{\rm p}^2a^2}\equiv{v_{\rm T}/\omega_{\rm p}a}$.
This feature simplifies the analysis of the behavior of the rate.
Plots of the function $\varepsilon_{\rm L}$ calculated 
numerically from equations (\ref{(B.2)}), (\ref{(B.8)}), and (\ref{(B.9)}) are given in Fig. 2.
In all cases, the low-temperature behavior of the dielectric 
constant is linear with respect to $x\sim v_{\rm T}$, i.e. the simple law 
(\ref{cold}) is recovered with
$v_{\rm T}$ in place of $v_F$ and with the dimensionless function 
$C(\omega)= C_0\omega_{\rm p}^2/\omega^2$, where $C_0$ waries between 1 and 2
depending on the boundary conditions\footnote{It should 
be noticed, however, that such linear behavior is not universal. 
For example, in the `triangular' 1D well, $U(x)\sim \vert x\vert$,
the dielectric constant is exponentially small at low temperatures, so that
the law (\ref{cold}) is never realized.
The dependence of the dielectric constant on the shape
of the self-consistent potential will be investigated elsewhere.}
(see insert in Fig. 2). 
In the opposite case of high temperatures, the behavior of the dielectric 
constant is different for specular and diffuse reflection.
However, realistic experimental parameters (see the estimates below) 
usually correspond to the limit $x\ll 1$, when the result is practically 
insensitive to the boundary conditions.
The behavior of the dielectric constant versus  temperature
may be easily understood from the resonance
condition (\ref{omega}).
Indeed, the oscillation frequency of a particle in the rectangular well
increases with increasing energy as $\sqrt\epsilon$ [see (\ref{(B.1)})].
This means that the upper `resonant' level has the energy 
$m_{\rm e}\omega^2a^2/2\pi^2$ [corresponding to $k=0$ in the resonance
condition (\ref{omega})];
all other `resonant' energy levels are at lower energies, 
near the bottom of the well.
As a result, for low temperature, when only the lowest energy levels
are populated, the number of `resonant' levels participating
in the absorption increases with increasing temperature as $T^2$.
Since the contribution of each level to the sum (\ref{(B.2)}) decreases as 
$T^{-3/2}$, the dielectric constant is proportional
to $v_{\rm T}$.
In the opposite case, for high temperature, the main contribution is given
by the single upper level with $\Omega(\epsilon)=\omega$.
Therefore, the rate of energy absorption is determined mainly by the derivative
of the distribution function in the vicinity of this level, which decreases like
$T^{-3/2}\sim v_{\rm T}^{-3}$ with increasing temperature.

Collisionless damping may influence  reflection and absorption of
a laser pulse by a film. 
For a thin film the reflection coefficient is small, 
$R\sim(a/\lambda)^2\ll 1$, and depends weakly on the imaginary part of the 
dielectric constant.
Therefore, the effect of  Landau damping on reflection is tiny.
We will not consider it here.
In contrast, 
the absorption coefficient is proportional 
to the small parameter $a/\lambda$ and depends strongly on the
imaginary part of the dielectric constant.
To estimate the relative contribution of Landau damping to absorption
we consider a thin dielectric film irradiated by an infrared or
optical femtosecond laser with intensity $\sim 10^{14} {\rm W/cm}^2$.
Laser pulses with such parameters are widely using in studies of the 
optical breakdown in dielectrics (see \cite{rudolph} and references therein).
The laser field produces inner ionization from the valence to the conduction band,
where electrons behave as a classical plasma with  temperature
close to the ponderomotive energy in the field (\ref{E}).

As follows from equation (\ref{Q3}), the energy absorbed per electron in the 
conduction band due to Landau damping is 
\beq
q_{\rm L}={\omega\tau E_0^2\over 8\pi n_{\rm e}}p^2\sin^2\vartheta{\varepsilon_{\rm L}\over
\vert\varepsilon\vert^2},
\label{qL}
\eeq
where $\tau$ is the duration of the laser pulse.
For $T_{\rm e}\simeq 10$ eV, $a\simeq 10$ nm and $\tau\simeq 100$ fs, from equation (\ref{(B.2)}) we have 
$x\approx 0.1$ and $\varepsilon_{\rm L}\approx 0.1$.
The electron density in the conduction band usually does not exceed 
the critical value corresponding to the plasma resonance in a dielectric,
when $\omega_{\rm p}=\omega\sqrt{\varepsilon_0^{\prime}+1}$.
For $1+\varepsilon_0^{\prime}\simeq 2\dots 3$ \cite{rudolph}, this
corresponds to $n_{\rm e}\simeq 10^{21}\ {\rm cm}^{-3}$.
When the density approaches  this critical value breakdown sets in.
Close to the critical density 
we obtain from equation (\ref{qL}) $q_{\rm L}\simeq 100$ eV.
This value corresponds to an effective relaxation time for  
Landau damping of $\tau_{\rm L}\simeq 10$ fs.
This is two orders of magnitude less than the electron-phonon relaxation
time, which is known to be around 1 ps for such parameters.
Another mechanism of energy absorption is inner ionization. It contributes 
about 10 eV per electron at the same intensity.
However, the energy transfer per electron due to Landau damping
decreases linearly with increasing film thickness $a$, while the
efficiency of absorption due to inner ionization practically does not
depend on the thickness, at least so long as it smaller than the skin depth.
Therefore, for films with thickness around 100 nm the contributions from Landau 
damping and inner ionization to the energy absorption will be comparable.
Obviously, with increasing thickness the effect of 
Landau damping vanishes.

The main conclusion of the estimations presented above is that for thin films
with a thickness below a few hundred nm 
Landau damping may dominate over the other mechanisms 
of energy absorption. 
As a result, the mean electron energy observed in spectra from
laser-irradiated films, which is approximately equal to the mean 
energy (\ref{qL}) absorbed from the field, should decrease proportionally 
to $a^{-1}$ with increasing thickness $a$.
Indeed, a mean electron energy decrease
with increasing thickness was clearly observed in experiments 
with thin carbon films \cite{gord}.

The expressions obtained above are valid, strictly speaking, for a 
monocrystalline film.
The consequences of Landau damping in polycrystalline films may be different
if the electrons remain trapped inside their respective crystal layers
under the action of the laser field.
This depends on the relation between the amplitude $x_0$ of the electron 
oscillations and the size $d$ of the crystals.
If $x_0\ll d$ then different crystals may be considered as independent 
systems.
In this case, the expression for the rate must be averaged over the 
orientations of the crystals so that it will become independent of 
the laser polarization.
Besides, spatial averaging will replace
the width $a$ of the film by the average size $d$ of the crystals.
Obviously, this limit is realized for low laser intensity 
(depending on $d$).
In the opposite case of high intensities, when $x_0\gg d$, a 
polycrystalline film is expected to give the same response as 
a monocrystalline film.

\ack
We are grateful to Dr. B.M.~Karnakov for useful advice.
This work was supported by the Deutsche Forschungsgemeinschaft (project 
no.436 RUS 113/676/0-1(R)) and the Russian Foundation for Basic Research.
Ph.A.K. and S.V.P.  also acknowledge support by  the  
Foundation of Noncommercial Programs "Dynasty".

\appendix
\section{Derivation of expression (\ref{main_1d}) for the absorption rate}

In the $(xy)$ plane, the electron motion is free.
Therefore, transitions between different energy levels  only occur due to the action of the
$z$ component of the electric field (\ref{E}).
The perturbation operator has the form
\beq
H_{\rm int}(z,t)={e\over m_{\rm e}c}p_zA_z(t),~~~~
A_z(t)=-c\int\limits_0^t{\mathcal{E}}_z(\tau)\rmd\tau.
\label{(A.1)}
\eeq
To first order in the laser field, we find
\beq
w_{\pm}={\pi e^2{\mathcal{E}}_z^2\over 2\hbar m_{\rm e}^2\omega^2}
\vert\langle n_1\vert p_z
\vert n\rangle\vert^2\delta(\epsilon_{n_1}-\epsilon_n\pm\hbar\omega).
\label{(A.2)}
\eeq
According to the selection rules for dipole transitions, the matrix 
elements of the operator (\ref{(A.1)}) are nonzero for levels with 
opposite parity, so that 
$$
n_1-n=2k+1,~~~~k=0,\pm 1,\pm 2,...
\nonumber
$$
In the semiclassical limit, we may replace the quantum numbers $n$ and $n_1$ by the
corresponding energies using the Bohr-Sommerfeld rule and the relations 
\beq
{\rmd n\over \rmd\epsilon}={1\over\hbar\Omega(\epsilon)},~~~
\epsilon_{n_1}-\epsilon_n=\hbar\Omega(\epsilon_n)(2k+1),~~~
\rho(\epsilon)=\hbar\Omega(\epsilon)f(\epsilon),
\label{(A.3)}
\eeq
which can be easily derived from the former. 
Here $\Omega(\epsilon)=2\pi/T(\epsilon)$ is the oscillation frequency of 
an electron with energy $\epsilon$ in the potential $U(z)$
and $f(\epsilon)$ is the classical distribution function, 
normalized to $\int \rmd\epsilon f(\epsilon)=1$.
Taking into account that $w_+(n,n_1)=w_-(n_1,n)$, we may rewrite (\ref{Q1})
in the form
\beq
\overline{Q}=\hbar\omega\sum_{n,n_1}w^-_{n,n_1}
[\rho(\epsilon_n)-\rho(\epsilon_{n_1})].
\label{(A.4)}
\eeq
Replacing the sum over the initial state with quantum number $n$ by an integral according to
$$
\sum_n\to\int \rmd n=\int {\rmd\epsilon\over\hbar\Omega(\epsilon)}
\nonumber
$$
and performing  this integration with the help of the $\delta$ function 
in equation (\ref{(A.2)}), then 
assuming that the typical electron energy is much higher than the photon
energy $\hbar\omega$, we obtain from equations (\ref{(A.2)})--(\ref{(A.4)})
\beq
\overline{Q}=-{\pi n_{\rm e}ae^2{\mathcal{E}}_z^2\over 2m_{\rm e}^2\omega}\sum_{n_1}
{\vert\langle n_1\vert p_z\vert n(n_1)\rangle\vert^2\over
\vert\Omega^{\prime}_{\epsilon}(\epsilon(n_1))\vert}
{\rmd\over \rmd\epsilon}\rho(\epsilon)_{\epsilon=\epsilon(n_1)},
\label{(A.5)}
\eeq
where $n(n_1)$ and $\epsilon(n_1)$ are functions of $n_1$ owing to the $\delta$ function
in the transition probability (\ref{(A.2)}).
Finally, we introduce a new summation index  $k$ instead of $n_1$
according to the second relation in (\ref{(A.3)}) [the resonance condition (\ref{omega})] 
and replace the matrix element
by its semiclassical limit.
For 1D systems, the semiclassical limit of the matrix element
is just the Fourier component of the corresponding classical quantity \cite{dau3}:
\beq
\langle n+s\vert p_z\vert n\rangle\to
p_s(\epsilon)={\Omega(\epsilon)\over 2\pi}\int\limits_0^{T(\epsilon)}
p_z(\epsilon,t)\rme^{\rmi s\Omega(\epsilon)t}\rmd t
\label{(A.6)}
\eeq
where $p_z(\epsilon,t)$ is the classical time-dependent momentum of the 
electron with the energy $\epsilon$.
Substituting equation (\ref{(A.6)}) into (\ref{(A.5)}) we end up with the rate (\ref{main_1d})
of energy absorption.

\section{The infinitely deep rectangular
well with specular and diffuse boundary conditions}

\subsection{Specular reflection}
For an infinitely deep one-dimensional rectangular well with width 
$a$, we have from equations (\ref{omega}), (\ref{Z}), and (\ref{(A.6)}):
\begin{eqnarray}
\Omega(\epsilon)&=&{\pi\over a}\sqrt{2\epsilon\over m_{\rm e}},~~~~~
\epsilon_k={m_{\rm e}\omega^2a^2\over 2\pi^2(2k+1)^2},
\nonumber \\
\vert p_s(\epsilon)\vert^2&=&{8m_{\rm e}\epsilon\over\pi^2s^2},~~~~~
Z_0(T)=a\sqrt{m_{\rm e}T\over 2\pi},
\label{(B.1)}
\end{eqnarray}
so that equation (\ref{epsL}) gives 
(recall that the superscript $s$ refers to specular reflection)
\beq
\varepsilon_{\rm L}^s=\bigg({\omega_{\rm p}\over\omega}\bigg)^2
{128\over\pi^6x^3}
\sum_{k=0}^{\infty}{\rme^{-4/\pi^3(2k+1)^2x^2}
\over(2k+1)^5},~~~x=\sqrt{8T\over\pi m_{\rm e}\omega^2a^2}
\equiv{v_{\rm T}\over\omega a}.
\label{(B.2)}
\eeq
For a relatively cold plasma in a wide well, we have $x\ll 1$, and
equation (\ref{(B.2)}) reproduces the law (\ref{cold}) with $v_{\rm T}$ 
in place of $v_F$ and $C=2\omega_p^2/\omega^2$.
In the opposite limit of a narrow well and a hot plasma, $x\gg 1$, and 
\beq
\varepsilon_{\rm L}^s\approx 4.13\bigg({\omega_{\rm p}\over\omega}\bigg)^2
{1\over x^3}.
\label{(B.3)}
\eeq
Let us reminde that the expression (\ref{(B.2)}) as well as the 
rate (\ref{(A.5)}) are valid in classical limit, when 
the condition $T\gg\hbar\omega$ is satisfied.
It is remarkable that for a deep rectangular well this restriction can be 
avoided if $(x\ll 1)$.
In this case the dielectric constant may be obtained directly from the rate
(\ref{(A.4)}). 
Using the exact expression for the matrix element
\beq
\langle n_1\vert p_z\vert n\rangle=-4\rmi{\hbar\over a}{nn_1\over{n_1^2-n^2}},
\label{(B.4)}
\eeq
replacing both sums by integrals over energies (this is possible only if
$x\ll 1$; in the general case only one sum can be replaced by an integral)
and taking into account that
$\epsilon(n)=\pi^2\hbar^2n^2/2m_{\rm e}a^2$, we obtain instead of (B.2)
\beq
\varepsilon_{\rm L}^s=2\bigg({\omega_{\rm p}\over\omega}\bigg)^2x
\sinh\eta K_1(\eta),~~~~~~
\eta={\hbar\omega\over 2T}.
\label{(B.5)}
\eeq
Here $K_1(\eta)$ is the McDonald function.
The areas of applicability of (\ref{(B.5)}) and (\ref{(B.2)}) overlap:
in the classical limit when $\eta\to 0$, 
the product $\sinh\eta K_1(\eta) \to 1$ and (\ref{(B.5)})
gives the low-temperature limit of (\ref{(B.2)}).

Beyond the linear approximation, the sinusoidally driven particle in the
infinitely deep well can serve as a paradigm of chaotic motion \cite{fuka}. The
breakdown of the invariant Kolmogorov-Arnold-Moser tori for increasing
laser field strength has been invoked to explain laser-induced damage thresholds \cite{bm87}.

\subsection{Diffuse scattering}
In a rectangular well Eq.~(\ref{diffv}) with the field (\ref{E}) 
has the solution
\begin{eqnarray}
v_+(t,\alpha)&=&v_0-{eE_0\over m_{\rm e}\omega}(\sin(\omega t+\alpha)-\sin\alpha),
\nonumber\\
x_+(t)&=&(v_0+{eE_0\over m_{\rm e}\omega}\sin\alpha)t
+{eE_0\over m_{\rm e}\omega^2}(\cos(\omega t+\alpha)-\cos\alpha),
\label{(B.6)}
\end{eqnarray}
where the subscript ``+'' indicates that the trajectory is calculated
for an electron that starts its motion at the left side of the well at $t=0$ 
(i.e. $x_+(0)=0$) with velocity $v(0)=v_0=\sqrt{2\epsilon/m_{\rm e}}$
when the field has the phase $\alpha$.
If the electron starts its motion from the right side, $x_-(0)=a$, 
then the trajectory is
\begin{eqnarray}
v_-(t,\alpha)=-v_0-{eE_0\over m_{\rm e}\omega}(\sin(\omega t+\alpha)-\sin\alpha),\nonumber\\
x_-(t)=a+(-v_0+{eE_0\over m_{\rm e}\omega}\sin\alpha)t
+{eE_0\over m_{\rm e}\omega^2}(\cos(\omega t+\alpha)-\cos\alpha).
\label{(B.7)}
\end{eqnarray}
The travel times $\tau_{\pm}(v_0,\alpha)$ are determined by the conditions 
$x_+(\tau_+)=a$ and $x_-(\tau_-)=0$. To first order in the field they are
\beq
\tau_{\pm}={a\over v_0}\bigg(1\mp{eE_0\over m_{\rm e}\omega^2a}\bigg\lbrack
\cos(z+\alpha)-\cos\alpha+z\sin\alpha\bigg\rbrack\bigg),~~
z={\omega a\over v_0}.
\label{(B.8)}
\eeq
Then the phase-averaged work (\ref{diffQe}) is equal to
\beq
A(\epsilon,\tau)=\tau\sqrt{2\epsilon\over m_{\rm e}}{e^2E_0^2\over\omega^2a}
\bigg\lbrack 2(1-\cos z)-z\sin z\bigg\rbrack.
\label{(B.9)}
\eeq
Finally, after averaging over the Maxwellian distribution 
we get instead of (B.2)
\beq
\varepsilon_{\rm L}^d={8\omega_{\rm p}^2\over\pi\omega^2 x}
\int\limits_0^{\infty}{\rmd z\over z^3}
\bigg\lbrack 2(1-\cos z)-z\sin z\bigg\rbrack
\rme^{-4/\pi x^2z^2},
\label{(B.10)}
\eeq
where the parameter $x$ is the same as in (\ref{(B.2)}).
We can see that in the case of diffuse reflection it is not necessary for 
the electron to satisfy the resonance condition (\ref{omega}) in order
to absorb energy: all electrons gain some net energy from the field.
However, the main contribution into absorption comes from the vicinities 
of the points $z=\pi(2k+1)$.
Using the definition of the variable $z$ in equation (\ref{(B.8)}) it is easy to find that this 
condition is equivalent to the resonance condition (\ref{omega}).

Finally, this result can be easily generalized to the case when the 
electron experiences exactly one specular reflection before 
the diffuse one.
Physically, this  corresponds to a thin film with different
boundary conditions on different surfaces, say to a film having a sublayer.
Using the same approach we get 
\begin{equation}
\varepsilon_{\rm L}^{s-d}={8\omega_{\rm p}^2\over\pi\omega^2 x}
\int\limits_0^{\infty}{\rmd z\over z^3}
\bigg\lbrack 1-\cos z-z\sin z\bigg\rbrack(1-\cos z)
\rme^{-4/\pi x^2z^2}.
\label{(B.11)}
\eeq

\begin{figure}
\centerline{\includegraphics[width=10.cm,clip=true]{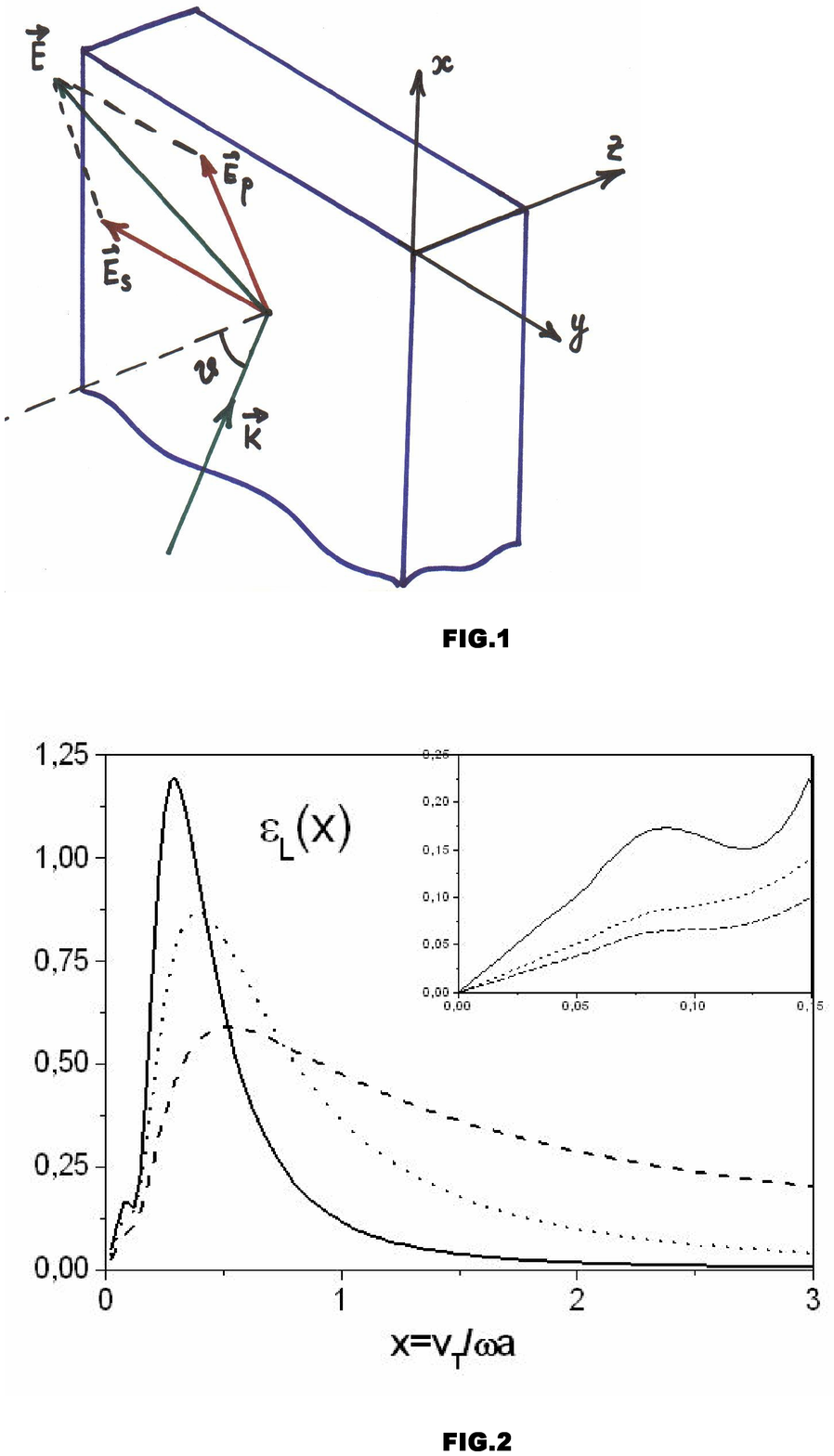}}
\caption{}
\label{fig}
\end{figure}

Both for diffuse and for specular-diffuse scattering, in the limit of low
temperatures ($x\ll 1$)
the dielectric constant $\varepsilon_{\rm L}$ goes to zero linearly with $x$ (see Fig. 2), regardless of the respective boundary condition.
In the opposite limit of high temperatures ($x\gg 1$), it decreases, too, but diffuse reflection leads to 
a substantially lower decrease, proportional to $v_{\rm T}^{-1}$, 
in comparison with specular scattering.


\vskip 2cm
\centerline{\bf{Figure captions}}

\vskip 1cm
{\bf FIG.1}
The geometry of the problem. The vector $\vecE_0$ is the electric field 
of the incident laser wave. It is decomposed into its s- and p-polarized
components, so that $\vecE_0=\vecE_s+\vecE_{\rm p}$, where 
the vector $\vecE_s$ is parallel to the $y$ axis, while  
the vector $\vecE_{\rm p}$ lies in the plane $(\vecn\veck)$, where $\veck$ is the 
wave vector of the incident laser field and $\vecn$ the unit normal vector
perpendicular to the plane of film. We use the notation $s=E_s/E_0$, $p=E_{\rm p}/E_0$, and $\vartheta$ is the angle between $\veck$ and $\vecn$.

\vskip 1cm
{\bf FIG.2} 
The imaginary part of the dielectric constant $\epsilon_{\rm L}$ as a function
of the dimensionless parameter $x=v_{\rm T}/\omega a$
calculated at $\omega=\omega_{\rm p}$ for the 1D rectangular well 
with specular (solid line), diffuse (dashed line), and specular-diffuse 
(dotted line) boundary conditions, respectively.
The behavior of the dielectric constant at low temperatures $(x\ll 1)$
is shown in the inset.

\end{document}